\documentclass[12pt, prd, showpacs]{revtex4}
\usepackage{amssymb}
\usepackage{amsmath}

\input{tcilatex}

\begin{document}

\title{Rotation as an origin of high energy particle collisions}
\author{O. B. Zaslavskii}
\affiliation{Department of Physics and Technology, Kharkov V.N. Karazin National
University, 4 Svoboda Square, Kharkov 61022, Ukraine}
\affiliation{Institute of Mathematics and Mechanics, Kazan Federal University, 18
Kremlyovskaya St., Kazan 420008, Russia}
\email{zaslav@ukr.net }

\begin{abstract}
We consider collision of two particles in rotating spacetimes without
horizons. If the metric coefficient responsible for rotation of spacetime is
big enough, the energy of collisions in the centre of mass frame can be as
large as one likes. This can happen in the ergoregion only. The results are
model-independent and apply both to relativistic stars and wormholes.
\end{abstract}

\maketitle

\section{Introduction}

In recent years, interest to high energetic phenomena in the strong
gravitational field arose after an important observation made in \cite{ban}.
It turned out that if two particles moving towards a black hole collide near
the horizon, their energy in the centre of mass frame can grow unbound. This
is so-called Ba\~{n}ados-Silk-West (BSW) effect. It renewed an interest to
more early works on high energy collisions \cite{pir1} - \cite{pir3} of
particles that moved in opposite directions or on circular orbits. Other
mechanisms of high energy collisions were also suggested. Some of them
implied that the event horizon either \ exists or is about to form.
Meanwhile, there are also quite different types of collisions not connected
with the horizon at all. One of examples is collision inside the ergosphere
(generally speaking, not in he vicinity of the horizon) when one of
particles has a large negative angular momentum (but modest individual
energy) \cite{gpergo}, \cite{mergo}. Quite recently, one more example of
high energy collision where the horizon is irrelevant was found. It was
shown \cite{wh1} that head-on collision of particles near the throat of the
Teo wormhole \cite{teo} leads to high $E_{c.m.}$

The goal of the present paper is to show that in space-times with rotation
rapid enough, high $E_{c.m.}$ can be obtained (i) both for wormholes and
relativistic stars and not only for head-on collisions but also for
particles moving in the same directions or orbiting a body or the wormhole
throat, (ii) it is an universal phenomenon irrespective of the details of
concrete models. The present work extends the list of mechanisms that give
rise to high energy particle collisions.

\section{Basic formulas}

We consider the stationary axially-symmetric metric

\begin{equation}
ds^{2}=-N^{2}dt^{2}+g_{\phi }(d\phi -\omega dt)^{2}+\frac{dr^{2}}{A}%
+g_{\theta }d\theta ^{2}\text{,}
\end{equation}%
where the coefficients do not depend on $t$ and $\phi $. Correspondingly,
the energy $E=-mu_{t}$ and the angular momentum $L=mu_{\phi }$ of a particle
moving in this background are conserved. Here, $m$ is its mass, $u^{\mu }$
being the four-velocity. These quantities are related by a simple formula 
\cite{k}%
\begin{equation}
X=\mathcal{E}N\text{,}  \label{cons}
\end{equation}%
where%
\begin{equation}
X=E-\omega L\text{,}  \label{xe}
\end{equation}%
\begin{equation}
\mathcal{E}=\frac{m}{\sqrt{1-V^{2}}}  \label{em}
\end{equation}%
has the meaning of the energy measured by the zero angular momentum observer 
\cite{72}, $V$ is the local speed measured by this observer.

The equations of motion of a test particle read%
\begin{equation}
m\frac{dt}{d\tau }=\frac{X}{N^{2}}\text{,}  \label{t}
\end{equation}%
\begin{equation}
m\frac{d\phi }{d\tau }=\frac{L}{g_{\phi }}+\frac{\omega X}{N^{2}}\text{,}
\label{phi}
\end{equation}%
\begin{equation}
m\frac{N}{\sqrt{A}}\frac{dr}{d\tau }=\sigma Z\text{,}  \label{r}
\end{equation}%
where $\sigma =\pm 1$ determines the sign of the radial momentum, 
\begin{equation}
Z=\sqrt{X^{2}-N^{2}(\frac{L^{2}}{g_{\phi }}+m^{2})}\text{.}  \label{Z}
\end{equation}

Let two particles 1 and 2 collide. Then, one finds for the energy in the
centre of mass frame $E_{c.m.}^{2}=-(m_{1}u_{1}^{\mu }+m_{2}u_{2}^{\mu
})(m_{1}u_{1\mu }+m_{2}u_{2\mu })$ that%
\begin{equation}
E_{c.m.}^{2}=m_{1}^{2}+m_{2}^{2}+2m_{1}m_{2}\gamma \text{,}  \label{cm}
\end{equation}%
where $\gamma =-u_{1\mu }u_{2}^{\mu }$ is the Lorentz factor of relative
motion. For simplicity, we restrict ourselves by motion in the equatorial
plane $\theta =\frac{\pi }{2}$. It follows from the equations of motion (\ref%
{t}) - (\ref{r}) that 
\begin{equation}
m_{1}m_{2}\gamma =\frac{X_{1}X_{2}-\sigma _{1}\sigma _{2}Z_{1}Z_{2}}{N^{2}}-%
\frac{L_{1}L_{2}}{g_{\phi }}\,\text{.}  \label{ga}
\end{equation}

We assume the forward in time condition $\frac{dt}{d\tau }\geq 0$. Then, (%
\ref{t}) entails 
\begin{equation}
X\geq 0\text{.}  \label{ft}
\end{equation}%
We discuss the situation when there is no horizon, so in the point of
collision $N=O(1)$. In what follows, we assume that%
\begin{equation}
\omega =\varepsilon \omega _{0}(r)>0\text{,}
\end{equation}%
where $\omega _{0}$ is the bounded function and the dimensionless parameter $%
\varepsilon \gg 1$. We are interested in the individual energies $E=O(1)$.
Then, it is seen from (\ref{ft}) that $L\leq 0$. It is worth noting that
independently of the sign of $L$, inside the ergoregion $\frac{d\phi }{d\tau 
}>0$.

\section{High energy collisions}

We consider three cases separately. It is essential that, if $L\neq 0$, $%
Z\approx X\approx \varepsilon \omega _{0}\left\vert L\right\vert $ is big.

\subsection{Case 1. Head-on collision, $\protect\sigma _{1}\protect\sigma %
_{2}=-1$.}

Hereafter, we assume that, although $g_{\phi }$ itself can be small (as it
takes place for the Teo wormhole \cite{teo}), it satisfies the inequality%
\begin{equation}
\frac{1}{\omega ^{2}g_{\phi }}\ll 1.  \label{cond}
\end{equation}%
Then, the main contribution to (\ref{ga}) comes from the terms with $\omega
. $If $L_{1}\neq 0$, it is seen from (\ref{cm}), (\ref{ga}) that%
\begin{equation}
E_{c.m.}^{2}\approx \frac{4\omega ^{2}\left\vert L_{1}L_{2}\right\vert }{%
N^{2}}  \label{h}
\end{equation}%
can be always made as large as one likes.

If $L_{1}=0$,%
\begin{equation}
E_{c.m.}^{2}\approx \frac{2\omega \left\vert L_{2}\right\vert (E_{1}+Z_{1})}{%
N^{2}}.  \label{1}
\end{equation}

\subsection{Case 2. Circular motion}

Let at least one of particles move on a circular orbit, so $Z_{1}=0$. Then,%
\begin{equation}
E_{c.m.}^{2}\approx \frac{2\omega ^{2}\left\vert L_{1}L_{2}\right\vert }{%
N^{2}}
\end{equation}%
and we arrive at the same conclusion. If $L_{1}=0$,%
\begin{equation}
E_{c.m.}^{2}\approx \frac{2\omega \left\vert L_{2}\right\vert E_{1}}{N^{2}}.
\label{11}
\end{equation}

\subsection{Case 3. Motion in the same direction, $\protect\sigma _{1}%
\protect\sigma _{2}=+1$}

This case is the most interesting one since fine-tuning is mandatory here in
analogy with the BSW effect \cite{ban}. We require individual energies $%
E_{1,2}$ to be finite. Then, one can obtain from (\ref{ga}) that for both
nonzero angular momenta $\gamma $ is finite as well, so the effect under
discussion is absent.

Let us now assume \ that $L_{1}=0$ (this can be thought of as a counterpart
of the critical particle in the BSW effect \cite{ban}) and $L_{2}<0.$ Then,
it is easy to obtain from (\ref{ga}) that%
\begin{equation}
E_{c.m.}^{2}\approx \frac{2\omega \left\vert L_{2}\right\vert (X_{1}-Z_{1})}{%
N^{2}}
\end{equation}%
can be made as big as we want.

As far as the relative motion of two particles is concerned, cases 1 and 2
are analogues of particle collisions near a rotating black hole considered
in \cite{pir1} - \cite{pir3} whereas case 3 is a rather close counterpart of
the BSW effect \cite{ban}. The quantity $X_{1}$ is modest whereas $X_{2}\sim
\omega $. Therefore, it is seen from (\ref{cons}) - (\ref{em}) that
kinematically, case 3 represents collision between a slow particle 1 and
rapid particle 2 according to (\ref{xe}) in full analogy with the kinematics
of the BSW effect \cite{k}.

\subsection{Comparison to the Teo wormhole}

For the Teo wormhole, $\omega =\frac{2a}{r^{3}}$, and, for motion the
equatorial plane, $g_{\phi }=r^{2}$, $N=1$ \cite{teo}. If collision occurs
near the throat, $r=b$, \ eq. (\ref{h}) gives us%
\begin{equation}
E_{c.m.}^{2}\approx \frac{16a^{2}\left\vert L_{1}L_{2}\right\vert }{b^{6}}
\end{equation}%
that corresponds to eq. (3.8) of \cite{wh1}. Condition (\ref{cond}) gives us 
$\frac{b^{4}}{a^{2}}\ll 1$ in agreement with the assumption $b\ll \sqrt{a}$
made in Sec III B of \cite{wh1}.

\section{Behavior of geometry}

If $\omega \rightarrow \infty $, the curvature invariants, generally
speaking, diverge. The rate of their growth is model-dependent since the
parameter $\varepsilon \,\ $can enter different metric coefficients. For the
Teo wormhole, the scalar curvature (see eq. (B1) of \cite{wh1}) $R=O(b^{-3})$
if $r=b$ and $R=O(b^{-6})$ if $r\neq b$ but has the order $b$. One has also $%
\omega (b)=O(b^{-3})=O(\varepsilon ).$ According to (\ref{1}), (\ref{11}),
for high energy collision on the throat $r=b$ the energy in the centre of
mass frame behaves according to $E_{c.m.}^{2}=$ $O(R)$ in case $L_{1}=0$ and 
$E_{c.m.}^{2}=O(R^{2})$ in other cases. Nonetheless, as the growth of
curvature is pure classic and does not contain quantum parameters, it is
possible to achieve intermediate large energies within the classical region,
without entering the Planck scale. Say, one can have simultaneously $\frac{%
E_{c.m.}}{m}\gg 1$ and $R\ll \frac{1}{L^{2}}$, where $L$ is some prescribed
scale since both inequalities contain different parameters. We do not
discuss this issue further since it is strongly model-dependent, meanwhile
we would like to make emphasis on model-independent features.

\section{Discussion}

It is seen from eq. (\ref{ga}) that in general, roughly speaking, there are
three main sources of high $E_{c.m.}$: (i) small $N$, (ii) high negative $L$%
, (iii) high $\omega $. Option (i) is related to collisions near black holes 
\cite{ban} - \cite{pir3}. It applies also to the collisions in the absence
of the horizon, provided the parameters of the system correspond to the
threshold of its formation and collision occurs just near such a would-be
horizon \cite{pj}, \cite{nosing}. Option (ii) was suggested in \cite{gpergo}
and generalized in \cite{mergo}. In the present work option (iii) as a
generic mechanism was considered that closes the list of possibilities.

In the derivation of our formulas, we assumed that in (\ref{xe}), the
quantity $E$ is negligible, provided $L\neq 0$, so 
\begin{equation}
\omega \gg \left\vert \frac{E}{L}\right\vert \text{.}  \label{e}
\end{equation}

It is worth stressing that in (\ref{e}) we require that $\omega $ be large
as compared to characteristics of a particle. However, we do \textit{not}
impose here direct restrictions on the parameters of a relativistic object
as such, say on the ratio $\frac{J}{M^{2}}$, where $J$ is its angular
momentum, $M$ being its mass. Such parameters can enter (\ref{e})
indirectly, through the quantity $\omega $ but, anyway, inequality (\ref{e})
relies on the particle's energy and angular momentum.

Also, it follows from derivation that in (\ref{Z}) we consider terms $N^{2}$
as a small corrections. Thus in addition to (\ref{e}),%
\begin{equation}
\omega ^{2}\gg \frac{N^{2}}{g_{\phi }}\text{.}
\end{equation}%
Therefore, the metric component $g_{00}=-N^{2}+g_{\phi }\omega ^{2}>0$, so
the effect takes place in the ergoregion only.

\section{Conclusions and outlook}

Among possible types of scenarios leading to high energy collisions, we
filled an important gap. We showed that fast rotation by itself leads to the
possibility to gain large energy of collision and, in this sense, it
represents an universal phenomenon. As far as the issue of large $E_{c.m.}$
is concerned, there is no need in detailed investigation of equations of
motion in some particular metrics, the results are obtained in a
model-independent way. They apply equally to rotating wormholes (thus
generalizing observations made in \cite{wh1} for the Teo wormhole) and to
rotating stars. The latter fact is a completely new venue for such
collisions. One can hope that this can be useful for astrophysics since, in
contrast to \ wormhole case, no exotic matter is needed. Also, in the
situation under discussion, there are no subtleties connected with the
relativistic time dilation and small fluxes \cite{mc}, \cite{com} since now
in the point of collision $N=O(1)$. The rate of growth of $E_{c.m}$ ($\omega 
$ or $\omega ^{2}$) depends on the type of collision and the fact whether or
not $L_{1}=0$. In any case, $L_{2}<0$. This is valid in the entire region,
where $\omega $ is large.

In contrast to the BSW effect near rotating black holes where $\omega $ is
close to the angular velocity of a black hole, now $\omega $ is formally as
big as one likes, so it is the dragging effect of the space-time itself that
leads to high energy of collisions in an universal manner. The price paid
for this is the appearance of large curvatures. However, one can choose the
situation when the curvature scale is far from, say the Planck value but,
nonetheless, the energy of collision $E_{c.m}$ is high enough. The mechanism
under discussion works in the ergoregion only.

Astrophysical applications are beyond the scope of the present paper.
However, as a separate issue, it would be of interest to trace the influence
of this mechanism on possible instabilities of ergoregion in relativistic
stars \cite{star1}, \cite{star2}. Also, it would be interesting to consider
the phenomenon under discussion using a more physical example than the Teo
wormhole \cite{teo} that was written by hand, without solving Einstein
equaitons. In particular, one can take the rotating wormhole obtained as a
solution of field equations with the phantom scalar field \cite{sc}.

The next step should consist in considering scenarios of the collisional
Penrose process. For wormholes, this is expected to generalize the results
found for the Teo wormhole in \cite{wh2}. For relativistic stars, this will
be a completely new issue.

All this needs separate treatment.

\begin{acknowledgments}
This work was funded by the subsidy allocated to Kazan Federal University
for the state assignment in the sphere of scientific activities.
\end{acknowledgments}


\begin{thebibliography}{99}
\bibitem{ban} M. Ba\~{n}ados, J. Silk and S.M. West, Phys. Rev. Lett. 
\textbf{103} (2009) 111102 [arXiv:0909.0169].

\bibitem{pir1} T. Piran, J. Katz, and J. Shaham, Astrophys. J. \textbf{196},
L107 (1975).

\bibitem{pir2} T. Piran and J. Shaham, Astrophys. J. \textbf{214}, 268
(1977).

\bibitem{pir3} T. Piran and J. Shanam, Phys. Rev. D \textbf{16}, 1615 (1977).

\bibitem{gpergo} A. A. Grib and Yu. V. Pavlov, Europhys. Lett. \textbf{101},
20004 (2013) [arXiv:1301.0698].

\bibitem{mergo} O. B. Zaslavskii, Mod. Phys. Lett. A. Vol. \textbf{28}, No.
11 (2013) 1350037 [arXiv:1301.4699].

\bibitem{wh1} N. Tsukamoto and C. Bambi, Phys. Rev. D \textbf{91}, 084013
(2015) [arXiv:1411.5778].

\bibitem{teo} E. Teo, Phys. Rev. D \textbf{58}, 024014 (1998)
[arXiv:gr-qc/9803098].

\bibitem{k} O. B. Zaslavskii, Phys. Rev. D \textbf{84}, 024007 (2011)
[arXiv:1104.4802].

\bibitem{72} J. M. Bardeen, W. H. Press, and S. A. Teukolsky, Astrophys. J. 
\textbf{178}, 347 (1972).

\bibitem{pj} M. Patil and P. S. Joshi, Phys. Rev. D \textbf{86}, 044040
(2012) [arXiv:1203.1803].

\bibitem{nosing} O. B. Zaslavskii, Phys. Rev. D \textbf{88}, 044030 (2013)
[arXiv:1305.6136].

\bibitem{mc} S. T. McWilliams, Phys. Rev. Lett. \textbf{110}, 011102 (2013)
[arXiv:1212.1235].

\bibitem{com} O. B. Zaslavskii, Phys. Rev. Lett. \textbf{111}, 079001 (2013)
[arXiv:1301.3429].

\bibitem{star1} V. Cardoso, P. Pani, M. Cadoni and M. Cavaglia, Phys. Rev. D 
\textbf{77}, 124044 (2008) [arXiv:0709.0532].

\bibitem{star2} V. Cardoso, P. Pani, M. Cadoni and M. Cavaglia, Class.
Quantum Grav. 25, 195010 (2008) [arXiv:0808.1615].

\bibitem{sc} B. Kleihaus and J. Kunz, Phys. Rev. D \textbf{90}, 121503(R)
(2014) [arXiv:1409.1503].

\bibitem{wh2} N. Tsukamoto and C. Bambi, Phys. Rev. D \textbf{91}, 104040
(2015) [arXiv:1503.06386].
\end{thebibliography}
\end{document}